\documentclass[5p,times]{elsarticle}
\usepackage{amsmath, amsthm, amssymb}
\usepackage{lineno,hyperref}
\usepackage{graphicx}
\usepackage{natbib}
\usepackage{color}
\usepackage{bm}

\journal{Matter and Radiation at Extremes}

\begin{document}

\begin{frontmatter}

\title{\textbf{\Large{A novel superconducting magnetic levitation method to support the laser fusion capsule by using permanent magnets}}}

\author{Xiaojia Li$^{a,*}$, Tingting Xiao$^a$, Fengwei Chen$^a$, Yingjuan Zhang$^a$, Weidong Wu$^{a,b,c,**}$}
\address{$^a$Research Center of Laser Fusion, China Academy of Engineering Physics, Mianyang 621900, China}
\address{$^b$Science and Technology on Plasma Physics Laboratory, Mianyang 621900, China}
\address{$^c$IFSA Collaborative Innovation Center, Shanghai Jiao Tong University, Shanghai 200240, China}

\cortext[cor1]{corresponding author}
\fntext[fn1]{shakalee@pku.edu.cn}
\cortext[cor2]{corresponding author}
\fntext[fn2]{wuweidongding@163.com}

\begin{abstract}
A novel magnetic levitation support method is proposed, which can relieve the perturbation caused by traditional support methods and provide more accurate position control of the capsule. This method can keep the perfect symmetry of the octahedral spherical hohlraum and has the characteristics in stability, tunability and simplicity. It is also favorable that all the results, such as supporting forces acting on the superconducting capsule, are calculated analytically, and numerical simulations are performed to verify these results. A typical realistic design is proposed and discussed in detail. The superconducting coating material is suggested, and the required superconducting properties are listed. Damped oscillation of the floating capsule in thin helium gas is discussed, and the restoring time is estimated.\\
\\
\textit{PACS codes:} 52.57.Fg; 74.70.Ad; 74.78.-W
\end{abstract}
\begin{keyword}
ICF capsule support, Magnetic levitation, Symmetry
\end{keyword}

\end{frontmatter}


\section{\label{sec1}Introduction}
Symmetry plays a crucial role in inertial confinement fusion (ICF) experiments, because any perturbation to the spherical symmetry will be strongly amplified through ablative hydrodynamic instabilities\cite{instability1,instability2}. To improve the radiation symmetry, indirect drive has been developed decades ago, spherical hohlraum has been reintroduced in recent years\cite{sphericalplasma2014,sphericalPRL2016,sphericalMRE2016}.
Capsule support is another important link in ICF experiments, which influences the symmetry around the capsule directly. Traditionally, people use thin wires, tents, low density foams or their hybrid systems\cite{wirefoam} to support the deuterium-tritium fuel capsule. However, in all the traditional methods, the supporting materials require direct contact with the capsule. Experimental results of National Ignition Facility (NIF) show that large perturbation exists where the tent departs from the capsule\cite{wirefoam}. Researchers are trying to make the wires thinner and the foam density lower, sacrificing their mechanical strength, but perturbation to the spherical symmetry is still inevitable, no matter how small it could be. It is then natural to think that only non-contact method can hopefully solve this problem and preserve the spherical symmetry around the capsule maximally. Different kinds of levitation methods has been pursued since the early years of ICF research since the 1970s\cite{earlyref1976}. To our knowledge, the idea of using superconductors to support ICF capsule was first presented by David Glocker\cite{earlyref1}, whose design was to support a permanently magnetized fuel pellet within a spherical superconducting shell. According to Meissner effect, superconductors are diamagnets, so that they will feel a repulsive magnetic force which points at the direction of the magnetic induction gradient. This magnetic force could be used to counteract gravity with proper design of the magnetic field. However, Glocker's design can no longer cooperate with nowadays ICF experiment. Y. Ishigaki et al. developed an accurate position control system for superconducting ICF capsule using three electromagnet coils\cite{earlyref2,earlyref3}. But technically, it does not seem possible to put three coils around the capsule without interfering with the laser paths, especially in the octahedral spherical hohlraums scheme where there are six laser entrance holes (LEHs)\cite{sphericalPRL2016}. There are also relative ideas such as using negative feedback digital circuits to control the capsule position\cite{earlyref1993}, using magnetic levitation to build a capsule transport system\cite{earlyref2014}, which we do not intend to comment too much on.

With symmetry preserved naturally, there are several issues we need to consider during schematic designing, such as stability, tunability and simplicity. First, the floating pellet must be in stable equilibrium in all degrees of freedom, both translational and rotational. Second, the equilibrium point should be tunable in a certain range and can be accurately controlled, because there are always slight mass differences between different pellets. Third, the supporting system should be compact enough to avoid all the laser paths and cooperate with other accessories around the hohlraum. Meanwhile, it should also be simple enough and easy to be installed. In this paper, we introduce a novel non-contact support method, in which the capsule is coated with superconducting thin film, and a potential well is designed right at the center of the hohlraum. Instead of coils\cite{earlyref2,earlyref3}, we use the mature Nd-Fe-B permanent magnets to generate the magnetic field. The typical magnetic flux density at the surface of a single Nd-Fe-B magnet is about $1.2$ tesla, then we have from Ampere's law that the corresponding effective surface current density is as high as $M\sim 10^6$A/m, so that the magnets could be much smaller than coils due to their intense magnetization. The cryogenic cooling temperature of the capsule is about $20$ Kelvin, giving us strict restriction in choosing material of the superconducting film. Another restriction comes from the ablation process, which demands that only low-Z atoms are acceptable\cite{lowZ}. $MgB_2$ becomes the ideal choice due to the above restrictions, whose average Z-value is relatively low and superconducting critical temperature is around $39$ Kelvin\cite{MgB2nature}. According to our initial settings, the film-coated pellet is supposed to behave like a solid perfect diamagnet. $MgB_2$ is a type II superconducting material\cite{MgB2film}, so that the pellet can be treated as a perfect diamagnetic sphere if and only if the outer magnetic flux density is below the lower critical field $B_{c1}$, and the film thickness is over several times the London penetration depth $\lambda$. Fortunately, the above conditions are not stringent. It is reviewed and concluded in Ref.\cite{MgB2film} that $B_{c1}$ is around $125$mT, and the corresponding $\lambda$ is about $40$nm. Meanwhile, our following calculations show that the required flux density in the vicinity of the capsule is only about $10$mT, and films of $100$nm to $200$nm are acceptable for the ablation\cite{ablation,thickness}. Taking these wonderful properties of $MgB_2$ into consideration, we believe that the $\textbf{E}-\textbf{J}$ complexities and flux pinning will not play a central role in the levitation process. The above discussions about $MgB_2$ have to be verified experimentally. While performing the simulations, we simply set the capsule's relative permeability $\mu_r=0$, and the capsule will behave similar enough to a superconductor (a perfect diamagnet). However, the levitating force may be weakened if the $MgB_2$ film is not as perfect as expected, so that we need to study this issue carefully in the future.

This article is organized as follows. In section \ref{sec2}, the elementary model is analyzed thoroughly. Using small pellet approximation, both near field and far field cases are discussed. The proposed supporting method is described in detail in section \ref{sec3}. In section \ref{sec4}, time needed for the capsule to restore equilibrium is estimated using perturbation method. Finally, we give a summary in section \ref{sec5}.

\section{\label{sec2}Analysis of the elementary model}
\begin{figure}
\begin{center}
  \includegraphics[width=7cm]{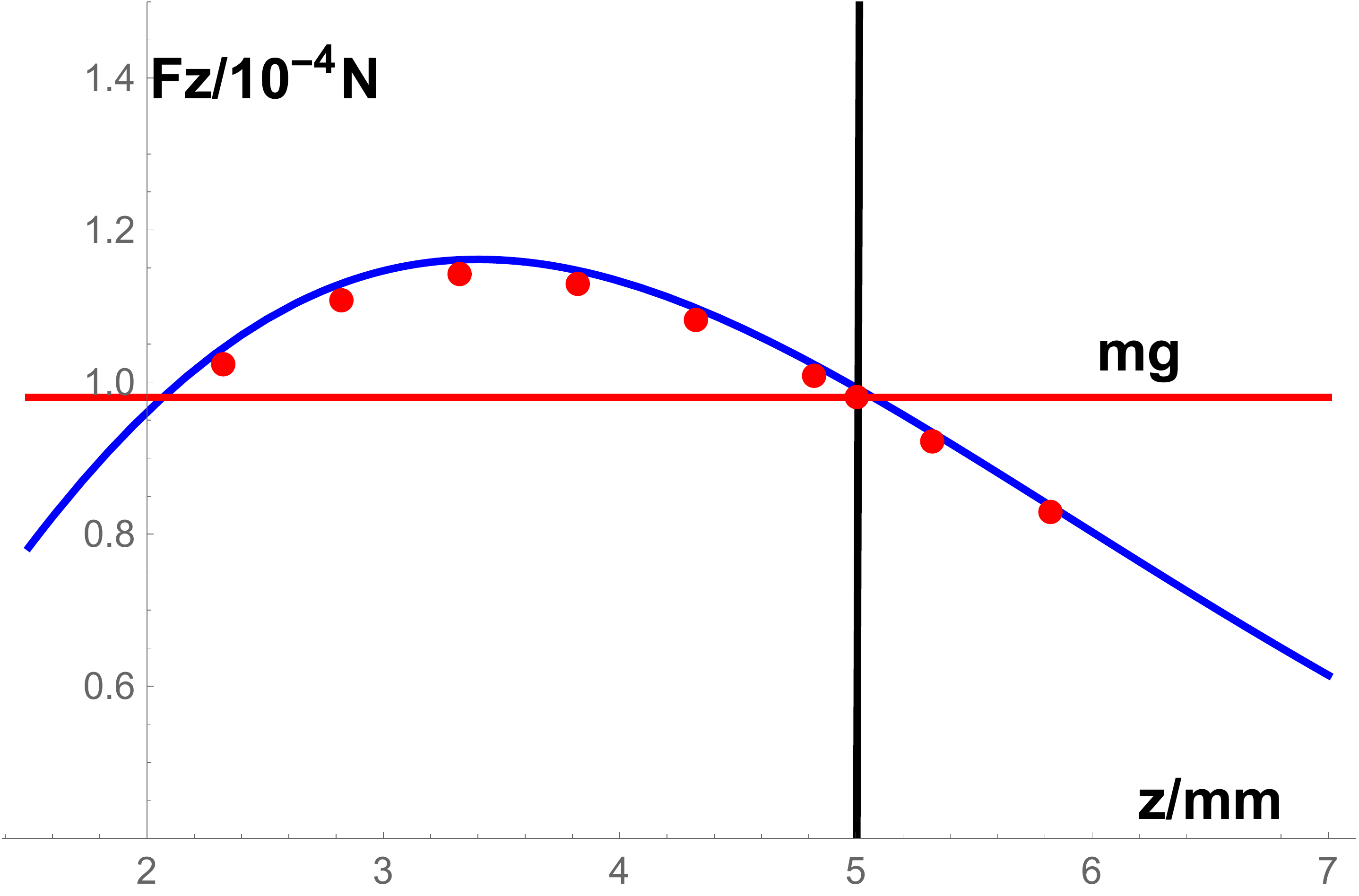}\\
  \caption{$F_z$ as a function of $z$ at $r=0$.} \label{Fz}
\end{center}
\end{figure}

\begin{figure}
\begin{center}
  \includegraphics[width=7cm]{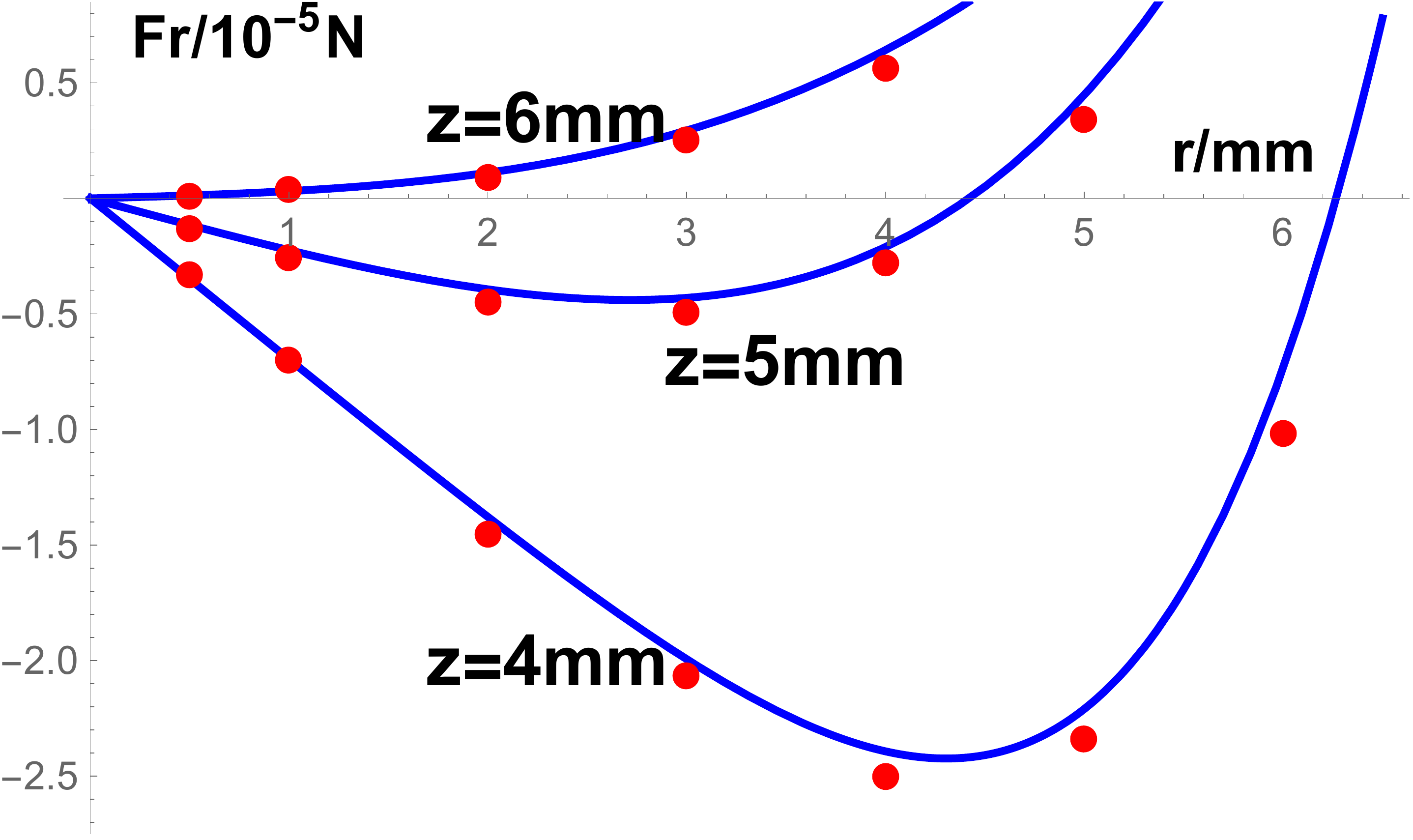}\\
  \caption{$F_r$ as a function of $r$ at different height.} \label{Fr}
\end{center}
\end{figure}

The simplest model in generating magnetic field must be a closed electric circle with radius $a$, carrying current $I$. As for permanent magnet, a uniformly magnetized wafer (a thin cylinder whose height $h$ is much smaller than the radius of undersurface $a$) is equivalent to such an electric circle, if its magnetization direction is parallel to its generatrix. This equivalence relation can be expressed quantitatively as
\begin{equation}\label{hIM}
I=h\cdot M.
\end{equation}
$M$ represents the norm of the magnetization vector $\textbf{M}$, similar notations will be widely used in the following text. Specifically, a thin cylindrical magnet with thickness $h=0.5$mm is equivalent to a circle with $500$ ampere current. We would keep $M=10^6$A/m a constant for simplicity. First, we need to calculate the magnetic field generated by the circle; second, we have to evaluate the magnetic force acting on the superconducting capsule. The above two steps are not difficult to work out separately, and most of the working procedures can be found in textbooks of electromagnetic dynamics. So we tend to list some useful results only, and focus our attention on the discussions and applications of these results. Suppose a circle (cylinder) is placed in the $z=0$ plane of the standard cylindrical coordinates, and its center is coincide with the original point. The current $I$ runs counter clockwise seeing from above. According to symmetry, the magnetic vector potential $\textbf{A}$ has only one non-zero component $A_{\phi}$, which can be written directly as:
\begin{equation}\label{magvector}
A_{\phi}(r,z)=\frac{\mu_0 I a}{4\pi}\int^{2\pi}_{0}\frac{cos\phi'}{\sqrt{r^2+z^2+a^2-2arcos\phi'}}d\phi'.
\end{equation}
The result of Eq. (\ref{magvector}) can be expressed by the secondary elliptic integration. These analytical calculations are performed using Wolfram Mathematica, and all figures except Fig.\ref{modelreal} are drawn by this software. After $\textbf{A}$ is known, we can derive the magnetic field by
$$\textbf{B} (r,z)=\bm{\nabla} \times \textbf{A}(r,z).$$
Vector $\textbf{B}$ has two non-zero components, $B_r$ and $B_z$. We do not introduce any approximations here, but in the next step, we have to introduce the so-called small pellet approximation, in order to calculate the magnetic forces perturbatively. First, assume that the pellet radius is much smaller than the curvature radius of the contours of the magnetic flux density, which allows us to ignore the field variance when deriving the magnetization of the pellet. Considering a superconducting sphere with radius $r_0$ placed into uniform magnetic field $\textbf{B}$, the sphere will be magnetized as a small dipole. Using the standard coordinates separation method in the spherical coordinates, together with proper boundary conditions, we can solve the magnetic scalar potential, which is expressed as the summation of two terms. The first term is just the original outer potential, and the second term represents a potential generated by a magnetic dipole:
\begin{equation}\label{m0}
\textbf{m}_\textbf{0}=-\frac{2\pi r_0^3}{\mu_0}\textbf{B}.
\end{equation}
This should be the dipole moment of the sphere itself, which is induced by the outer field. The second meaning of small pellet approximation is that we ignore the radius of the pellet and treat it as a point dipole placed in curved magnetic field, then the corresponding potential energy is:
\begin{equation}\label{Um}
U_m=-\frac{1}{2}\textbf{m}_\textbf{0}\cdot\textbf{B}.
\end{equation}
One may notice that there is an extra factor $1/2$ compared with the normal expression. This factor arises because half of the energy is used to magnetize the pellet. After these preparations we can finally get the expression of the magnetic force:
\begin{equation}\label{Fm}
\textbf{F}_\textbf{m}=-\bm{\nabla} U_m.
\end{equation}
We test the effectiveness of these equations by comparing the results with numerically simulated ones. Considering a near field case, if we set $a=9$mm, $r_0=0.8$mm and $I=355$A a priori, the pellet with $m=10$mg can be floated at the destination height of $z_0=5$mm, right above the center of the circle. Similarly, we set $h=0.355$mm while doing the simulation using finite element method software. The results are shown in Fig.\ref{Fz} and Fig.\ref{Fr}, in which the blue lines are calculated forces and the simulated forces are shown by red points.

Results obtained by different methods show good agreement with each other, which demonstrates that the small pellet approximation works excellent in such a near field example. We can see from Fig.\ref{Fz} that the simulated supporting force $F_z$ is always slightly smaller than the calculated one. This is a systematic error caused by the measurement of distance. Apparently, $F_z(z)$ is a convex function in the vicinity of the equilibrium point, so that the supporting force decreases faster than linear with the increase of $z$. But when performing the simulations, the distances are measured from the geometry centers of the objects.

\begin{figure}
\begin{center}
  \includegraphics[width=7cm]{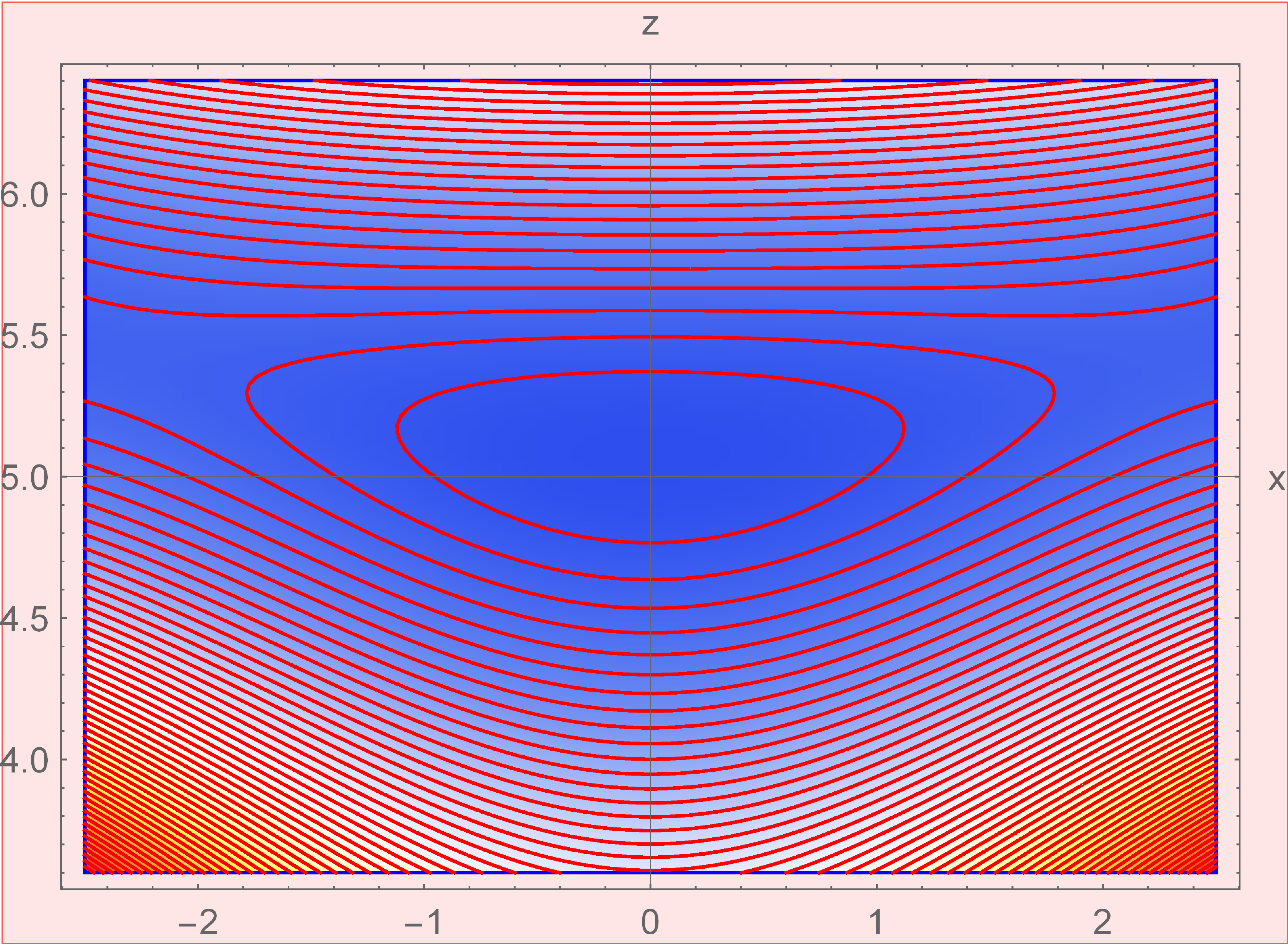}\\
  \caption{Distribution of total potential $U_t$. Red lines represent the contour surfaces of $U_t$}\label{Utfig}
\end{center}
\end{figure}

Now we discuss the stability problem we put forward before. Translational equilibrium is acquired naturally by putting the capsule into the equilibrium point $z=5$mm, $r=0$mm, where $F_z=mg$ and $F_r=0$. In the vicinity of the equilibrium point, stability in $z$ direction is obvious. In $r$ direction there is a critical height (approximately $z/a=0.63$), above which the equilibrium will be unstable because $F_r$ will become positive. The stable range can be found by defining the total potential energy:
\begin{equation}\label{Ut}
U_t=U_m+mgz,
\end{equation}
where $g=9.8$N/kg is the gravitational acceleration. The structure of $U_t$ is shown in Fig.\ref{Utfig}, which confirms the existence of the axially symmetric three dimensional potential well of about $4\textrm{mm}\times1\textrm{mm}$ in size. Such a well is large enough to trap the capsule.
In the rotational degrees of freedom, stability is naturally guaranteed by the magnetization mechanism of superconductors. According to Eq.(\ref{m0}), the direction of the dipole is just opposite to the outer field $\textbf{B}$, so that the rotational moment acting on the capsule is
\begin{equation}\label{L=mb}
\textbf{L}=\textbf{m}_\textbf{0}\times\textbf{B}
\propto\textbf{B}\times\textbf{B}=0,
\end{equation}
at any case.

We now turn to the far field case, which means the distance from the dipole to the circle $R$ is much greater than both $r_0$ and $a$. The small pellet approximation is then automatically satisfied. The circle then can also be treated as a magnetic dipole with moment $\textbf{m}_\textbf{c}=\pi r_0^2 h \textbf{M}$, whose magnetic scalar potential at a field point $\textbf{R}$ is already well known as:
\begin{equation}\label{farphi}
\varphi_m=\frac{\textbf{m}_\textbf{c}\cdot \textbf{R}}{4\pi R^3}.
\end{equation}
Eq.(\ref{farphi}) comes from the leading term of Taylor expansion of the exact expression of $\varphi_m$, which tells us that in far field case, all other components of the magnetic force $\textbf{F}$ become insignificant compared with $F_z$ along $z$ axis.

We briefly summarize this section as follows. The magnetic field of a thin cylindrical permanent magnet is studied both analytically and numerically. A stable equilibrium point exists in near field case, with proper adjustment of relevant parameters. In far field case, the magnet wafer acts as a rod, which provides only a force perpendicular to its undersurface. These conclusions will be used in the next section.

\section{\label{sec3}Realistic design of the supporting scheme}
In the six LEHs spherical hohlraums scheme\cite{sphericalPRL2016}, one LEH is set right under the capsule, making it impossible to put a solid magnet underneath. However, a thin cylinder with a coaxial hole is doable if the radius of the hole is large enough. Such a gasket like magnet is equivalent to two electric loops placed at the inner and outer circle of the magnet. The two loops share the same current intensity but opposite current directions. We can see from common sense that near field distribution is dominated mainly by the inner loop, but the outer loop will weaken the field and the supporting force, making the stable equilibrium range lower than the elementary one-circle case. Up to some specific height, the magnetic field will switch direction. This critical height is determined by the ratio of the inner and outer loop diameters. Consequently, we can not increase the levitating height by simply increasing the effective current. Fortunately, the gasket like magnet is placed out of the spherical hohlraum and the existence of the hole allows us to put the magnet higher, but the outer diameter must be set smaller at the same time to avoid laser paths. To achieve satisfactory destination height and stability, small wafers of permanent magnets can be used as rods to counteract part of the capsule gravity and refine the shape of the potential well. These small wafers can be placed on the outer surface of the hohlraum.
\begin{figure}
\begin{center}
  \includegraphics[width=7cm]{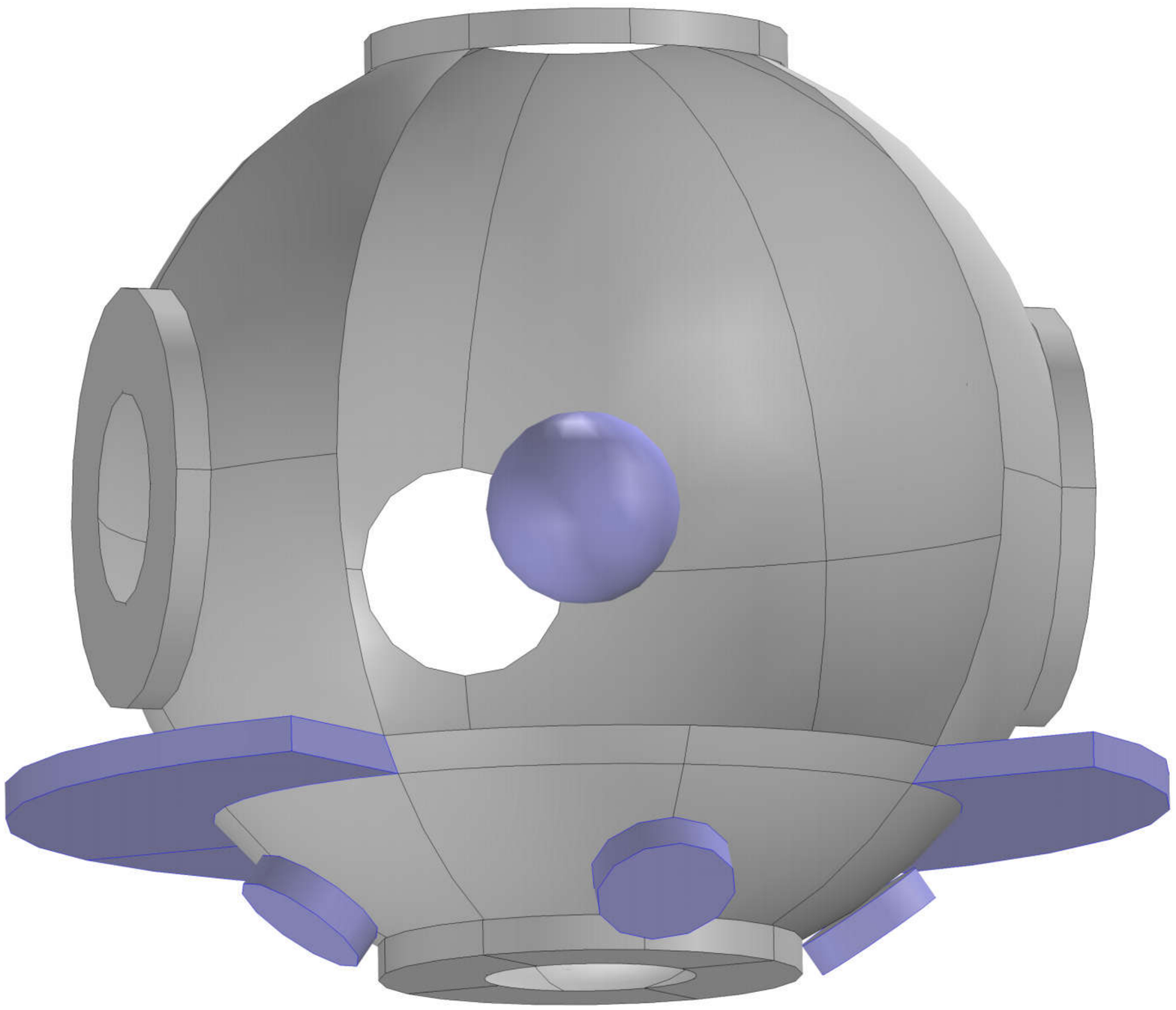}\\
  \caption{Typical design of the magnets' placements.}\label{modelreal}
\end{center}
\end{figure}

\begin{figure}
\begin{center}
  \includegraphics[width=7cm]{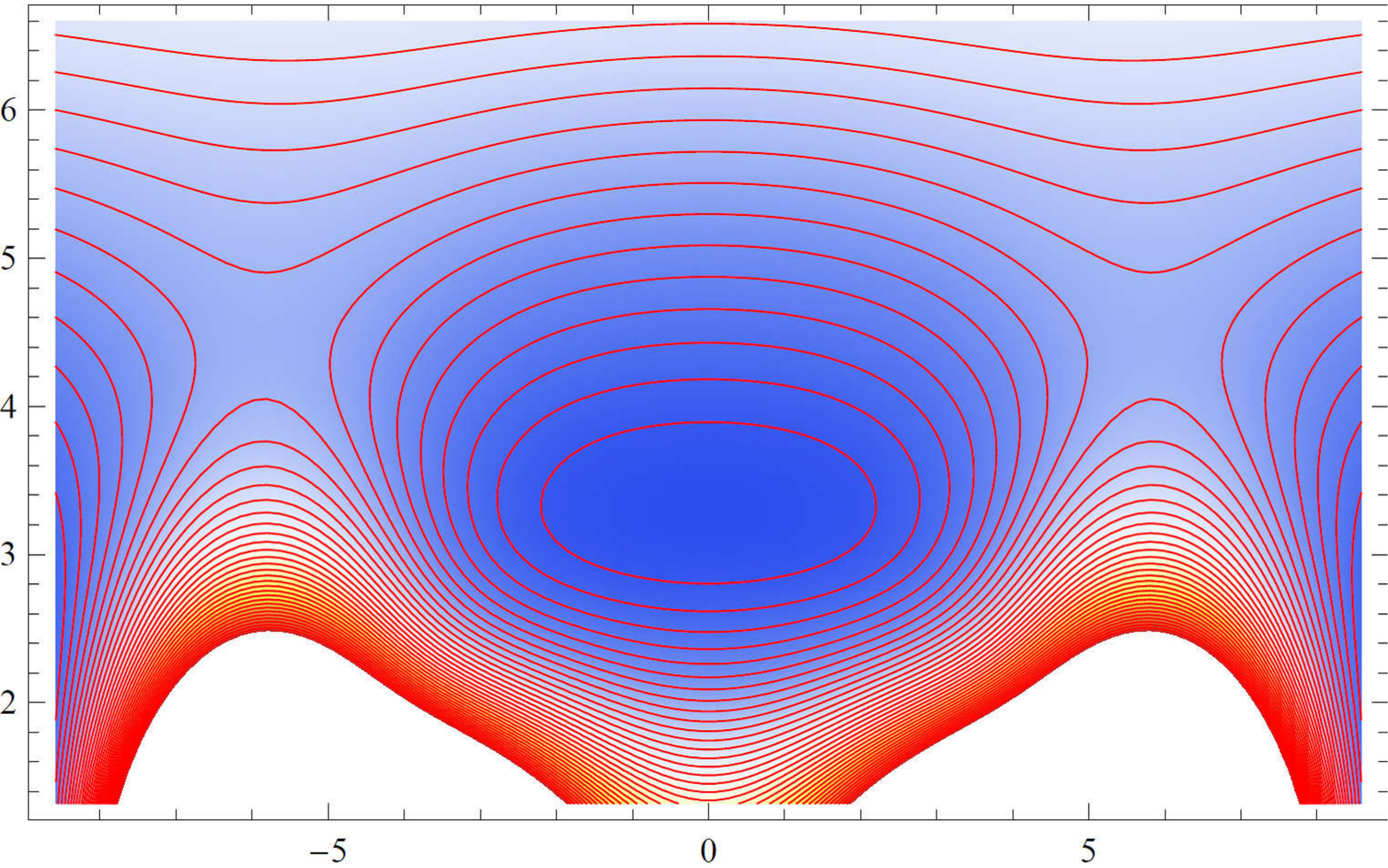}\\
  \caption{Cut plane (x-z plane) of the total potential well.}\label{Utxzplane}
\end{center}
\end{figure}
We demonstrate the effectiveness of this scheme by proposing a detailed design with typical geometry parameters as follows. We put the gasket like magnet on the $z=0$ plane as before, whose inner and outer radius is $r_{1}=4.4$mm and $r_{2}=6.6$mm respectively. A spherical hohlraum of $r_{3}=5.5$mm can be put tightly in the gasket, with its center at $z_0=3.3$mm. The destination height above the gasket is then $3.3$mm because the capsule must be put at the center of the hohlraum. Four small wafers with radius $r_4=0.88$mm are stick on the hohlraum under the gasket right below the $x$ and $y$ axes. Their undersurfaces are set $\theta=35$ degrees away from horizontal, facing the destination point, as shown in Fig.\ref{modelreal}. The thickness of the gasket and the wafers are $d_1=d_4=0.4$mm(with equivalent currents $400$A). If the radius of the LEHs is set to be $1.2$mm, the largest allowed laser entrance angle will be $\theta_i=67$ degrees, which is large enough for experiments\cite{sphericalMRE2016}. Without loss of generality, we set the wafers to be magnetized toward the capsule. To enhance (other than weaken) the field of the wafers, the gasket should be magnetized downward.

Using Eqs.(\ref{magvector})-(\ref{Fm}), together with some trivial translational and rotational coordinates transformations, we can obtain the supporting force for a $r_0=1.1$mm superconducting pellet is $F_z=2.093\times 10^{-4}$N. The simulated result is $2.092\times 10^{-4}$N, which is also slightly smaller as expected. After inputting the corresponding equilibrium mass ($m=21.36$mg) to Eq.(\ref{Ut}), we can plot the total potential of $x-z$ plane in Fig.\ref{Utxzplane}. A potential well of $10\textrm{mm}\times3.5\textrm{mm}$ is shown clearly, which should be satisfactory for practical application. The potential well on the $y=x$ cut plane looks similar but slightly larger, because the wafers are placed under the $x$ and $y$ axes, which breaks the axial symmetry of the original system.

With stability guaranteed by the potential well, tunability is another problem to tend to. This supporting system is highly flexible. Parameters as $r_1$, $d_1$, $r_4$, $d_4$ and $\theta$ can be tuned to adjust different hohlraum sizes and different capsule masses, which we can call coarse tuning. In traditional methods, the position error of the capsule is mainly caused by the supporting wires or foams, whose mechanical strength is limited by geometry. This problem no longer exists in our method because Nd-Fe-B magnets are solids with excellent mechanical strength. Accurate position control of the capsule can be achieved by fine tuning the position of the gasket through mechanical methods. As for simplicity, this system is compact enough to keep away from all laser paths.

\section{\label{sec4}Estimation of restoring time of the floating capsule in thin helium}
This section is relatively independent, but essential for the application of this supporting method. Recall that our initial purpose is to fix the capsule to a exact point with minimal error. The potential well ensures the specific equilibrium point, but it is difficult to find this very point at beginning. Suppose the capsule is released statically at some random point within $1$mm from the equilibrium point, then the capsule will become a harmonic oscillator approximately. Taking the actual situation into consideration that the hohlraum is filled with helium gas, the amplitude of the oscillation must decrease with time. This system is just a typical three dimensional damped oscillator. How much time it takes to achieve equilibrium is one thing that engineers care about. Generally, the equation of motion for a one dimensional damped oscillator can be written as:
\begin{equation}\label{oscillator}
q''(t)+\frac{\alpha}{m}q'(t)+\frac{\kappa}{m} q(t)=0.
\end{equation}
In our case, $\alpha=6\pi r_0 \eta$ is the friction coefficient caused by helium viscosity, and $\kappa=-F(q)/q$ is the linear restoring coefficient determined by the total potential $U_t$. Before solving the above equation, we evaluate the order of magnitude of both terms. From material data base we find that at the typical temperature ($T=20$K) and pressure ($p=0.1$atm) of the hohlraum, helium viscosity $\eta=3.55\times10^{-6}\textrm{Pa}\cdot\textrm{s}$, leading to $\alpha/m=3.44\times10^{-3}\textrm{s}^{-1}$. Meanwhile, even at the weakest direction of the potential well, $\kappa/m\sim 700\textrm{s}^{-2}$. That is to say, the weak damping condition is perfectly satisfied, which allows us to solve Eq.(\ref{oscillator}) perturbatively. Taking the second term as a perturbation, together with the initial values $q(0)=q_0\sim 1\textrm{mm}$ and $q'(0)=0$, the solution up to $O(\frac{\alpha}{m})$ is then written as:
\begin{equation}\label{solution}
q(t)=q_0e^{-\frac{t}{t_0}}\cos(\sqrt{\frac{\kappa}{m}}t),
\end{equation}
where we defined the eigen time $t_0=2m/\alpha=580\textrm{s}$. We can see from Eq.(\ref{solution}) that the oscillation magnitude decreases exponentially with time, and will be one order of magnitude smaller after every ($\ln10\cdot t_0=22$) minutes. For the realistic three dimensional case, the position error is magnified only by a factor of $\sqrt{3}$. We conclude that the position error of the capsule should be smaller than $10\mu\textrm{m}$ within $50$ minutes, which is already more accurate than traditional methods. This estimation is rather conservative, because the weak damping approximation no longer works if the amplitude is small enough. As a result, the real restoring time should be much shorter. In addition, the fill tube also plays a role in damping the oscillation, even additional damping mechanism could be introduced if necessary. As far as we know, this result is acceptable for experiments.


\section{\label{sec5}Summary}
In this paper, we propose a novel non-contact method to support the capsule in octahedral spherical hohlraum, in which only superconducting film and permanent magnets are needed. The feasibility of the method is theoretically verified by both analytical and numerical methods. The high symmetry around the capsule can be maximally preserved because of the non-contact nature of this method. Since there is only one stable equilibrium point in the potential well, and the position of the magnets can be tuned more precisely, the capsule position can be controlled more accurately. This system is adequately small in geometry size, thus applicable in engineering. In the cylindrical hohlraum case, even the wafers are not necessary because the gasket is almost free in $z$ direction. Time required to restore equilibrium is estimated conservatively, which we believe is not a problem to worry about.

The main challenge of this project might be the fabrication of the high quality $MgB_2$ thin film on a small sphere, which affects the diamagnetism property of the capsule directly. Corresponding experimental research is being carried out simultaneously by our team.


\end{document}